\documentclass{llncs} 
\usepackage{amsfonts}
\usepackage{amsmath}
\usepackage{amssymb} 
\usepackage{pstricks} 
\usepackage{complexity}
\usepackage{epsfig}
\usepackage[normalem]{ulem}
\usepackage{array}
\usepackage{pstricks,pst-plot} %

\newcommand{\tnb}[0]{\ensuremath{b\:}}
\newcommand{\tnbUline}[0]{\ensuremath{\underline{b}\:}}
\newcommand{\tnOne}[0]{\ensuremath{1\,}}
\newcommand{\tnOneUline}[0]{\ensuremath{\underline{1}\,}}
\newcommand{\tnZero}[0]{\ensuremath{0\,}}
\newcommand{\tnZeroUline}[0]{\ensuremath{\underline{0}\,}}

\newcommand{\tnMarkZero}{\ensuremath{0\hspace{-1.2ex}/\,}} 
\newcommand{\tnMarkZeroUline}{\ensuremath{\underline{0\hspace{-1.2ex}/}}\,} 

\newcommand{\tnMarkOne}{\ensuremath{1\hspace{-1.2ex}/\,}} 
\newcommand{\tnMarkOneUline}{\ensuremath{\underline{1\hspace{-1.2ex}/}}\,} 

\title{Small weakly universal Turing machines}
\author{Turlough Neary\inst{1} and Damien Woods\inst{2}}
\institute{TASS, Department of Computer Science,\\ National University of 
Ireland Maynooth, Ireland. \email{tneary@cs.may.ie} \and Department of Computer Science,\\ University 
College Cork, Ireland. \email{d.woods@cs.ucc.ie}}
\begin{document}
\maketitle
\begin{abstract}
We give small universal Turing machines with state-symbol pairs of $(6,2)$, $(3,3)$ and $(2,4)$. These machines are weakly universal, which means that they have an infinitely repeated word to the left of their input and another to the right. They simulate Rule 110 and are currently the smallest known weakly universal Turing machines.
\end{abstract}

\section{Introduction}\label{sec:Introduction}
Shannon~\cite{Shannon1956} was the first to consider the problem of finding the smallest universal Turing machine, where size is the number of states and symbols. Here we say that a Turing machine is standard if it has a single one-dimensional tape, one tape head, and is deterministic~\cite{HopcroftUllman1979}. Over the years, small universal programs were given for a number of variants on the standard model. By generalising the model we often find smaller universal programs. One variation on the standard model is to allow the blank portion of the Turing machine's tape to have an infinitely repeated word to the left, and another to the right. We refer to such universal machines as weakly universal Turing machines, and they the subject of this work.

Beginning in the early sixties Minsky and Watanabe engaged in a vigorous competition to see who could come up with the smallest universal Turing machine~\cite{Minsky1960,Minsky1962,Watanabe1960,Watanabe1961,Watanabe1972}. In 1961 Watanabe~\cite{Watanabe1961} gave a 6-state, 5-symbol universal Turing machine, the first weakly universal machine. In 1962, Minsky~\cite{Minsky1962} found a small 7-state, 4-symbol universal Turing machine. Not to be out-done, Watanabe improved on his earlier machine to give 5-state, 4-symbol and 7-state, 3-symbol weakly universal machines~\cite{Watanabe1972,Nozaki1969}.

\begin{figure}[t]
\begin{center}
\newcommand{\dwtnfigurefontsize}{\scriptsize}
\newcommand{\dwtnfigurelegendfontsize}{\scriptsize}
\psset{unit=3ex}
\begin{pspicture}(-5,-1.5)(22,19)

\psset{dotsize=4pt,dotstyle=o}
\psdot (10.2,16.2)
\put (11,16) {\dwtnfigurelegendfontsize : standard universal machine}

\psset{dotstyle=diamond*,linewidth=1pt}
\psdot (10.2,15.2)
\put (11,15) {\dwtnfigurelegendfontsize : semi-weakly universal machine}

\psset{dotstyle=square,dotsize=4pt}
\psdot (10.2,14.2)
\put (11,14) {\dwtnfigurelegendfontsize : weakly universal machine}

\psset{dotstyle=square*,dotsize=4pt}
\psdot (10.2,13.2)
\put (11,13) {\dwtnfigurelegendfontsize : new weakly universal machine}

\psset{linestyle=solid}
\psline (8,12.15)(10.4,12.15)
\put (11,12) {\dwtnfigurelegendfontsize : standard universal curve}

\psset{linestyle=dashed, dash=3pt 3pt}
\psline (8,11.15)(10.4,11.15)
\put (11,11) {\dwtnfigurelegendfontsize : weakly universal curve}

\psset{linestyle=solid}
{\tiny \psaxes[ticksize=1pt]{->}(20,19)}
\rput (4.5,-2) {states}
\put (-5,4) {symbols}


\psset{linestyle=solid}
\psline (2,18)(2,19)
\psline (2,18)(3,18)
\psline (3,9)(3,18)
\psline (3,9)(4,9)
\psline (4,6)(4,9)
\psline (4,6)(5,6)
\psline (5,5)(5,6)
\psline (5,5)(6,5)
\psline (6,4)(6,5)
\psline (6,4)(9,4)
\psline (9,3)(9,4)
\psline (9,3)(18,3)
\psline (18,2)(18,3)
\psline (18,2)(20,2)

\psset{linestyle=dashed, dash=3pt 3pt}
\psline (2,4)(2,19)
\psline (2,4)(3,4)
\psline (3,4)(3,3)
\psline (3,3)(6,3)
\psline (6,3)(6,2)
\psline (6,2)(20,2)

\psset{dotsize=4pt,dotstyle=o}
\psdot (2,18)
\psdot (3,9)
\psdot (4,6)
\psdot (5,5)
\psdot (6,4)
\psdot (9,3)
\psdot (18,2)

\psset{dotsize=4pt,linewidth=1pt} %
\psset{dotstyle=diamond*,linewidth=1pt}
\psdot (5,4)
\psdot (7,3)
\psdot (4,5)
\psdot (3,7)

\psset{dotsize=4pt,dotstyle=square}
\psdot (2,5)
\psdot (3,4)
\psdot (4,3)
\psdot (7,2)

\psset{dotstyle=square*,dotsize=4pt}
\psdot (2,4)
\psdot (3,3)
\psdot (6,2)

\end{pspicture}

\end{center}
\caption{State-symbol plot of small universal Turing machines. Each of our new weak machines is represented by a solid square. These machines induce a weakly universal curve.}\label{fig:states-sym-july-07}
\end{figure}
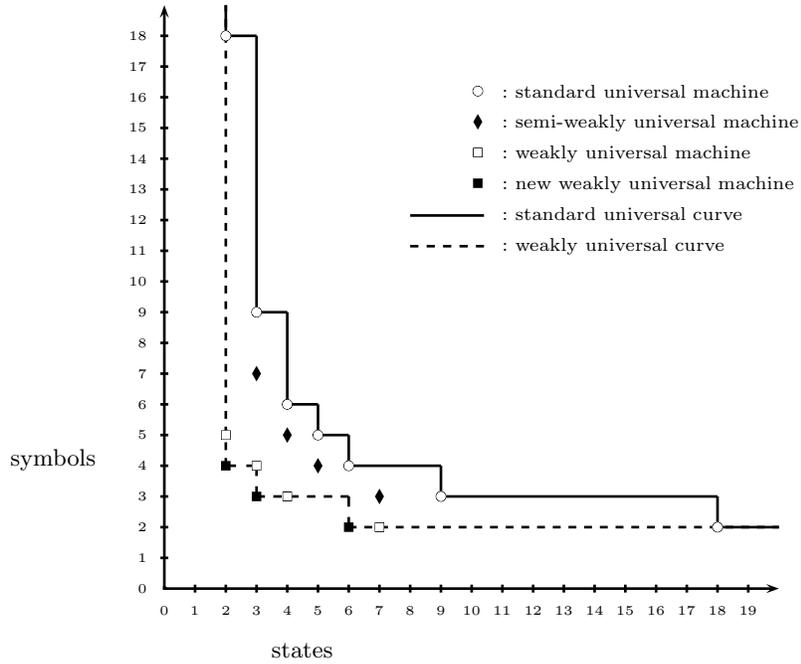

The 7-state universal Turing machine of Minsky has received much attention. Minsky's machine simulates Turing machines via 2-tag systems, which were proved universal by Cocke and Minsky~\cite{CockeMinsky1964}. The technique of simulating 2-tag systems, pioneered by Minsky, was extended by Rogozhin~\cite{Rogozhin1982} to give the (then) smallest known universal Turing machines for a number of state-symbol pairs. These 2-tag simulators were subsequently reduced in size by Rogozhin~\cite{Rogozhin1996}, Kudlek and Rogozhin~\cite{KudlekRogozhin2002}, and Baiocchi~\cite{Baiocchi2001}. Neary and Woods~\cite{NearyWoods2007} gave small universal machines that simulate Turing machines via a new variant of tag systems called bi-tag systems. Each of the smallest 2-tag or bi-tag simulators are plotted as circles in Figure~\ref{fig:states-sym-july-07}. These (standard) machines induce a universal curve.

The small weak machines of Watanabe have received little attention. In particular the 5-state and 7-state machines seem little known and are largely ignored in the literature. It is worth noting that while all other weak machines simulate Turing machine via other simple models, Watanabe's weak machines simulate Turing machines directly. His machines are the most time efficient of the small weak machines. More precisely, let $t$ be the running time of any deterministic single tape Turing machine $M$, then Watanabe's machines are the smallest weak machines that simulate $M$ with a time overhead of $O(t^2)$.

We often refer to Watanabe's machines as being semi-weak. Semi-weak machines have an infinitely repeated word to one side of their input, and on the other side they have a (standard) infinitely repeated blank symbol. Recently, Woods and Neary~\cite{WoodsNeary2007A} have given 3-state, 7-symbol and 4-state, 5-symbol semi-weakly universal machines that simulate cyclic tag systems. All of the smallest semi-weakly universal machines are given as diamonds in Figure~\ref{fig:states-sym-july-07}.

Cook~\cite{Cook2004} and Wolfram~\cite{Wolfram2002} recently gave weakly universal Turing machines, smaller than Watanabe's semi-weak machines, that simulate the universal cellular automata Rule 110. These machines have state-symbol pairs of $(7,2)$, $(4,3)$, $(3,4)$, $(2,5)$ and are plotted as hollow squares in Figure~\ref{fig:states-sym-july-07}.

Here we present weakly universal Turing machines with state-symbol pairs of $(6,2)$, $(3,3)$, $(2,4)$, making them the smallest known weakly universal machines. Our machines simulate (single tape, deterministic) Turing machines in time $O(t^4 \log^2 t)$, via Rule 110. These machines are plotted as solid squares in Figure~\ref{fig:states-sym-july-07} and induce a weakly universal curve. It is interesting to note from Figure~\ref{fig:states-sym-july-07} that the smallest universal machines, and the smallest semi-weakly universal machines, are both symmetric about the line where states equals symbols, whereas the smallest weakly universal machines are not.

Weakness has not been the only variation on the standard model in the search for small universal Turing machines. Priese~\cite{Priese1979} gave a 2-state, 4-symbol machine with a 2-dimensional tape, and a 2-state, 2-symbol machine with a 2-dimensional tape and 2 tape heads. Margenstern and Pavlotskaya~\cite{MargensternPavlotskaya2003} gave a 2-state, 3-symbol Turing machine that is universal when coupled with a finite automaton. This machine uses only 5 instructions. Margenstern and Pavlotskaya also show that the halting problem is decidable for machines of this type with 4 instructions. Their result implies that it is not possible to have a 4 instruction universal machine of this type, that simulates any Turing machine $M$ and halts if and only if $M$ halts. Hence they have given the smallest possible universal machine of this type.

For results relating to the time complexity of small universal Turing machines see~\cite{NearyWoods2006C,NearyWoods2006,WoodsNeary2006,WoodsNeary2007}.

\subsection{Preliminaries}
The Turing machines considered in this paper are deterministic and have a single bi-infinite tape. We let~$U_{m,n}$ denote our weakly universal Turing machine with~$m$ states and~$n$ symbols. We write $c_1\vdash c_2$ if a configuration~$c_2$ is obtained from~$c_1$ via a single computation step. We let $c_1\vdash^{s}c_2$ denote a sequence of~$s$ computation steps, and let $c_1\vdash^{\ast}c_2$ denote zero or more computation steps.

\section{Rule 110}\label{sec:Rule110}
Rule 110 is a very simple (2 state, nearest neighbour, one dimensional) cellular automaton. It is composed of a sequence of cells $\ldots p_{-1}p_{0}p_{1}\ldots$ where each cell has a binary state $p_i\in\{0,1\}$. At timestep $s+1$, the value 
$p_{i,s+1}=F(p_{i-1,s},p_{i,s},p_{i+1,s})$ of the cell at position $i$ is given by the synchronous local 
update function~$F$
\begin{align}\label{eq:Rule110}
\begin{split}
F(0,0,0)=0 \qquad & \qquad  F(1,0,0)=0 \\
F(0,0,1)=1 \qquad & \qquad F(1,0,1)=1 \\
F(0,1,0)=1 \qquad & \qquad F(1,1,0)=1 \\
F(0,1,1)=1 \qquad & \qquad F(1,1,1)=0
\end{split}
\end{align}

\begin{figure}[t]
\begin{center}
\psset{unit=2.5ex}
\begin{pspicture}(-9,-6)(30,2)
\psset{linestyle=solid}

\put(-7,1.4){\scriptsize $c_0$}
\put(-7,0.3){\scriptsize $c_1$}
\put(-7,-0.75){\scriptsize $c_2$}
\put(-7,-1.8){\scriptsize $c_3$}
\put(-6.8,-3.3){\scriptsize $\vdots$}

{\scriptsize
\put(0,2.3){\ldots}
\put(1.2,2.2){-9}
\put(2.2,2.2){-8}
\put(3.2,2.2){-7}
\put(4.2,2.2){-6}
\put(5.2,2.2){-5}
\put(6.2,2.2){-4}
\put(7.2,2.2){-3}
\put(8.2,2.2){-2}
\put(9.2,2.2){-1}
\put(10.3,2.2){0}
\put(11.3,2.2){1}
\put(12.3,2.2){2}
\put(13.3,2.2){3}
\put(14.3,2.2){4}
\put(15.3,2.2){5}
\put(16.3,2.2){6}
\put(17.3,2.2){7}
\put(18.3,2.2){8}
\put(19.3,2.2){9}
\put(20.3,2.3){\ldots}
}

\pspolygon[fillstyle=solid, fillcolor=black](-4,1)(-4,2)(-3,2)(-3,1)
\pspolygon[fillstyle=solid, fillcolor=black](-1,1)(-1,2)(1,2)(1,1)
\pspolygon[fillstyle=solid, fillcolor=black](2,1)(2,2)(7,2)(7,1)
\pspolygon[fillstyle=solid, fillcolor=black](10,1)(10,2)(11,2)(11,1)
\pspolygon[fillstyle=solid, fillcolor=black](13,1)(13,2)(15,2)(15,1)
\pspolygon[fillstyle=solid, fillcolor=black](16,1)(16,2)(21,2)(21,1)

\pspolygon[fillstyle=solid, fillcolor=black](-5,0)(-5,1)(-3,1)(-3,0)
\pspolygon[fillstyle=solid, fillcolor=black](-2,0)(-2,1)(3,1)(3,0)
\pspolygon[fillstyle=solid, fillcolor=black](6,0)(6,1)(7,1)(7,0)
\pspolygon[fillstyle=solid, fillcolor=black](9,0)(9,1)(11,1)(11,0)
\pspolygon[fillstyle=solid, fillcolor=black](12,0)(12,1)(17,1)(17,0)
\pspolygon[fillstyle=solid, fillcolor=black](20,0)(20,1)(21,1)(21,0)
\pspolygon[linecolor=gray](-1,0)(-1,1)(23,1)(23,0)

\pspolygon[fillstyle=solid, fillcolor=black](-5,-1)(-5,0)(-1,0)(-1,-1)
\pspolygon[fillstyle=solid, fillcolor=black](2,-1)(2,0)(3,0)(3,-1)
\pspolygon[fillstyle=solid, fillcolor=black](5,-1)(5,0)(7,0)(7,-1)
\pspolygon[fillstyle=solid, fillcolor=black](8,-1)(8,0)(13,0)(13,-1)
\pspolygon[fillstyle=solid, fillcolor=black](16,-1)(16,0)(17,0)(17,-1)
\pspolygon[fillstyle=solid, fillcolor=black](19,-1)(19,0)(21,0)(21,-1)
\pspolygon[fillstyle=solid, fillcolor=black](22,-1)(22,0)(23,0)(23,-1)
\pspolygon(-1,-1)(-1,0)(23,0)(23,-1)

\pspolygon[fillstyle=solid, fillcolor=black](-2,-2)(-2,-1)(-1,-1)(-1,-2)
\pspolygon[fillstyle=solid, fillcolor=black](1,-2)(1,-1)(3,-1)(3,-2)
\pspolygon[fillstyle=solid, fillcolor=black](4,-2)(4,-1)(9,-1)(9,-2)
\pspolygon[fillstyle=solid, fillcolor=black](12,-2)(12,-1)(13,-1)(13,-2)
\pspolygon[fillstyle=solid, fillcolor=black](15,-2)(15,-1)(17,-1)(17,-2)
\pspolygon[fillstyle=solid, fillcolor=black](18,-2)(18,-1)(23,-1)(23,-2)

\pspolygon[fillstyle=solid, fillcolor=black](-3,-3)(-3,-2)(-1,-2)(-1,-3)
\pspolygon[fillstyle=solid, fillcolor=black](0,-3)(0,-2)(5,-2)(5,-3)
\pspolygon[fillstyle=solid, fillcolor=black](8,-3)(8,-2)(9,-2)(9,-3)
\pspolygon[fillstyle=solid, fillcolor=black](11,-3)(11,-2)(13,-2)(13,-3)
\pspolygon[fillstyle=solid, fillcolor=black](14,-3)(14,-2)(19,-2)(19,-3)
\pspolygon[fillstyle=solid, fillcolor=black](22,-3)(22,-2)(23,-2)(23,-3)

\pspolygon[fillstyle=solid, fillcolor=black](-4,-4)(-4,-3)(-1,-3)(-1,-4)
\pspolygon[fillstyle=solid, fillcolor=black](-1,-4)(-1,-3)(1,-3)(1,-4)
\pspolygon[fillstyle=solid, fillcolor=black](4,-4)(4,-3)(5,-3)(5,-4)
\pspolygon[fillstyle=solid, fillcolor=black](7,-4)(7,-3)(9,-3)(9,-4)
\pspolygon[fillstyle=solid, fillcolor=black](10,-4)(10,-3)(15,-3)(15,-4)
\pspolygon[fillstyle=solid, fillcolor=black](18,-4)(18,-3)(19,-3)(19,-4)
\pspolygon[fillstyle=solid, fillcolor=black](21,-4)(21,-3)(23,-3)(23,-4)

\pspolygon[fillstyle=solid, fillcolor=black](-5,-5)(-5,-4)(-3,-4)(-3,-5)
\pspolygon[fillstyle=solid, fillcolor=black](0,-5)(0,-4)(1,-4)(1,-5)
\pspolygon[fillstyle=solid, fillcolor=black](3,-5)(3,-4)(5,-4)(5,-5)
\pspolygon[fillstyle=solid, fillcolor=black](6,-5)(6,-4)(11,-4)(11,-5)
\pspolygon[fillstyle=solid, fillcolor=black](14,-5)(14,-4)(15,-4)(15,-5)
\pspolygon[fillstyle=solid, fillcolor=black](17,-5)(17,-4)(19,-4)(19,-5)
\pspolygon[fillstyle=solid, fillcolor=black](20,-5)(20,-4)(23,-4)(23,-5)

\pspolygon[fillstyle=solid, fillcolor=black](-4,-6)(-4,-5)(-3,-5)(-3,-6)
\pspolygon[fillstyle=solid, fillcolor=black](-1,-6)(-1,-5)(1,-5)(1,-6)
\pspolygon[fillstyle=solid, fillcolor=black](2,-6)(2,-5)(7,-5)(7,-6)
\pspolygon[fillstyle=solid, fillcolor=black](10,-6)(10,-5)(11,-5)(11,-6)
\pspolygon[fillstyle=solid, fillcolor=black](13,-6)(13,-5)(15,-5)(15,-6)
\pspolygon[fillstyle=solid, fillcolor=black](16,-6)(16,-5)(21,-5)(21,-6)

\psline[linecolor=gray](-5,-6)(-5,2)
\psline[linecolor=gray](-4,-6)(-4,2)
\psline[linecolor=gray](-3,-6)(-3,2)
\psline[linecolor=gray](-2,-6)(-2,2)
\psline[linecolor=gray](-1,-6)(-1,2)
\psline[linecolor=gray](0,-6)(0,2)
\psline[linecolor=gray](1,-6)(1,2)
\psline[linecolor=gray](2,-6)(2,2)
\psline[linecolor=gray](3,-6)(3,2)
\psline[linecolor=gray](4,-6)(4,2)
\psline[linecolor=gray](5,-6)(5,2)
\psline[linecolor=gray](6,-6)(6,2)
\psline[linecolor=gray](7,-6)(7,2)
\psline[linecolor=gray](8,-6)(8,2)
\psline[linecolor=gray](9,-6)(9,2)
\psline[linecolor=gray](10,-6)(10,2)
\psline[linecolor=gray](11,-6)(11,2)
\psline[linecolor=gray](12,-6)(12,2)
\psline[linecolor=gray](13,-6)(13,2)
\psline[linecolor=gray](14,-6)(14,2)
\psline[linecolor=gray](15,-6)(15,2)
\psline[linecolor=gray](16,-6)(16,2)
\psline[linecolor=gray](17,-6)(17,2)
\psline[linecolor=gray](18,-6)(18,2)
\psline[linecolor=gray](19,-6)(19,2)
\psline[linecolor=gray](20,-6)(20,2)
\psline[linecolor=gray](21,-6)(21,2)
\psline[linecolor=gray](22,-6)(22,2)
\psline[linecolor=gray](23,-6)(23,2)

\psline[linecolor=gray](-5.5,2)(23.5,2)
\psline[linecolor=gray](-5.5,1)(23.5,1)
\psline[linecolor=gray](-5.5,0)(23.5,0)
\psline[linecolor=gray](-5.5,-1)(23.5,-1)
\psline[linecolor=gray](-5.5,-2)(23.5,-2)
\psline[linecolor=gray](-5.5,-3)(23.5,-3)
\psline[linecolor=gray](-5.5,-4)(23.5,-4)
\psline[linecolor=gray](-5.5,-5)(23.5,-5)
\psline[linecolor=gray](-5.5,-6)(23.5,-6)
\end{pspicture}\caption{Seven consecutive timesteps of Rule 110. These seven timesteps are applied to the background ether that is used in the proof~\cite{Cook2004} of universality of Rule 110. Each black or each white square represents, a Rule 110 cell containing, state 1 or 0 respectively. Each cell is identified by the index given above it. To the left of each row of cells there is a configuration label that identifies that row.} \label{fig:Rule110Steps}
\end{center}
\end{figure}
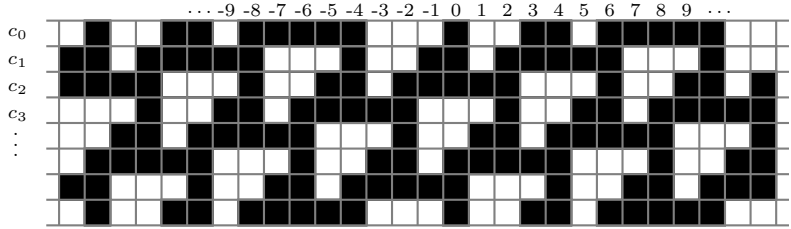
Rule 110 was proven universal by Cook~\cite{Cook2004} and Wolfram~\cite{Wolfram2002}. Recently, Neary and Woods~\cite{NearyWoods2006C,NearyWoods2006C1} proved that Rule 110 simulates Turing machines efficiently in polynomial time $O(t^3\log t)$, an exponential improvement. Note that, in order to calculate this upper bound we substitute space bounds for time bounds whenever possible in the analysis. It turns out that we can further improve the simulation time to $O(t^2 \log t)$ (this result is as yet unpublished). Rule 110 simulates cyclic tag systems in linear time. The weak machines in this paper, and in~\cite{Cook2004,Wolfram2002}, simulate Rule 110 with a quadratic polynomial increase in time and hence simulate Turing machines in time $O(t^4 \log^2 t)$. It is worth noting that the prediction problem~\cite{Greenlaw1995} for these machines is $\P$-complete, and this is also the case when we consider only bounded initial conditions~\cite{NearyWoods2006C}.

\section{Three small weakly universal Turing machines}\label{sec:SmallUTMs}
The following observation is one of the reasons for the improvement in size over previous machines~\cite{Cook2004,Wolfram2002}, and gives some insight into the simulation algorithm we use. Notice from Equation~\eqref{eq:Rule110} that the value of the update function $F$, with the exception of $F(0,1,1)$ and $F(1,1,1)$, may be determined using only the rightmost two states. Each of our universal Turing machines exploit this fact as follows. The machines scan from right to left, and in six of the eight cases they need only remember the cell immediately to the right of the current cell~$i$ in order to compute the update for $i$. Thus for these six cases we need only store a single cell value, rather than two values. The remaining two cases are simulated as follows. If two consecutive encoded states with value $1$ are read, it is assumed that there is another encoded $1$ to the left and the update $F(1,1,1)=0$ is simulated. If our assumption proves false (we instead read an encoded $0$), then our machine returns to the wrongly updated cell and simulates the update $F(0,1,1)=1$.

Before giving our three small Rule~110 simulators, we give some further background explanation. Rule 110 simulates Turing machines via cyclic tag systems. A Rule 110 instance that simulates a cyclic tag system computation is of the following form (for more details see~\cite{Cook2004,Wolfram2002}). The input to the cyclic tag system is encoded in a contiguous finite number of Rule 110 cells. On the left of the input a fixed constant word (representing the `ossifiers') is repeated infinitely many times. On the right, another fixed constant word (representing the cyclic tag system program/appendants, and the `leaders') is repeated infinitely many times. Both of these repeated words are independent of the input.

As in~\cite{Cook2004,Wolfram2002}, our weakly universal machines operate by traversing a finite amount of the tape from left to right and then from right to left. This simulates a single timestep of Rule 110 over a finite part of the encoded infinite Rule 110 instance. With each simulated timestep the length of a traversal increases. So that each traversal is of finite length, the left blank word $l$ and the right blank word $r$ of each of our weak machines must have a special form. These words contain special subwords or symbols that terminate each traversal, causing the tape head to turn. When the head is turning it `deletes' any symbols that caused a turn. Thus the number of cells that are being updated increases monotonically over time. This technique simulates Rule 110 properly if the initial condition is set up so that within each repeated blank word, the subword between each successive turn point is shifted one timestep forward in time.

In the sequel we describe the computation of our three machines by showing a simulation of the update on the ether in Figure~\ref{fig:Rule110Steps}. In the next paragraph below, we outline why this example is in fact general enough to prove universality. First, we must define blank words that are suitable for this example. The left blank word $l$, on the Turing machine tape, encodes the Rule 110 sequence $0001$. In the initial configuration as we move left each subsequent sequence $0001$ is one timestep further ahead. To see this note from Figure~\ref{fig:Rule110Steps} that $0001$ occupies, cells $-7$ to $-4$ in configuration $c_1$, cells $-11$ to $-8$ in $c_2$, cells $-15$ to $-12$ in $c_3$, etc. Similarly, the right blank word $r$ encodes the Rule 110 sequence $110011$. Looking at the initial configuration, as we move right from cell $0$, in the first blank word the first four cells $1100$ are shifted two timesteps ahead, and the next two cells $11$ are shifted a further one timestep. To see this note from Figure~\ref{fig:Rule110Steps} that $1100$ occupies cells $1$ to $4$ in $c_2$ and $11$ occupies cells $5$ and $6$ in $c_3$. In each subsequent sequence the first four cells $1100$ are shifted only one timestep ahead and the last two cells $11$ are shifted one further timestep. In each row the ether in Figure~\ref{fig:Rule110Steps} repeats every 14 cells and if the number of timesteps $s$ between two rows is $s\equiv 0\mod 7$ then the two rows are identical. The periodic nature of the ether, in both time and space, allows us to construct such blank words.

It should be noted that the machines we present here, and those in~\cite{Cook2004,Wolfram2002}, require suitable blank words to simulate a Rule 110 instance directly. If no suitable blank words can be found (i.e.~if they do not contain the specific subwords that we use to terminate traversals) then it may be the case that the particular instance can not be simulated directly. In the sequel our machines simulate the background ether that is used in the universality proof of Rule 110~\cite{Cook2004,Wolfram2002}. The gliders that move through this ether are periodic in time and space, and so we can construct blank words where the ether includes the subwords that terminate traversals. By this reasoning, our example is sufficiently general to prove that our machines simulate Turing machines via Rule 110 and we do not give a full (and possibly tedious) proof of correctness. For $U_{3,3}$ we explicitly simulate three updates from Figure~\ref{fig:Rule110Steps}, which is general enough so that an update [Equation~\eqref{eq:Rule110}] on each of the eight possible three state combinations is simulated.  We give shorter examples for the machines $U_{2,4}$ and $U_{6,2}$ as they use the same simulation algorithm as~$U_{3,3}$.

The machines we present here do not halt. Cook~\cite{Cook2004} shows how a special glider may be produced during the simulation of a Turing machine by Rule 110. This glider may be used to simulate halting as the encoding can be such that it is generated by Rule 110 if and only if the simulated machine halts. The glider would be encoded on the tape of our machines as a unique, constant word.

\subsection{$U_{3,3}$}\label{sec:U(3,3)}
We begin by describing an \emph{initial} configuration of $U_{3,3}$. To the left of, and including, the tape head position, the Rule~110 state $0$ is encoded by $0$, and the Rule~110 state $1$ is encoded by either $1$ or $b$. The word $1b0$ is used to terminate a left traversal. (Note an exception: the~$1$ in the subword $1b0$  encodes the Rule 110 state~$0$.) To the right of the tape head position, the Rule~110 state~$0$ is encoded by~$1$, and the Rule 110 state~$1$ is encoded by~$0$ or~$b$. The tape symbol~$0$ is used to terminate a right traversal. The left and right blank words, described in paragraph 4 of Section~\ref{sec:SmallUTMs}, are encoded as $\tnZero\tnZero\tnOne\tnb$ and $\tnZero\tnb\tnOne\tnOne\tnZero\tnb$ respectively.

\begin{table}[h]
\begin{center}
\begin{tabular}{c@{\;\;}|@{\;\;}c@{\;\;}c@{\;\;}c}
 & $u_{1}$ & $u_{2}$  & $u_{3}$  \\ \hline
$0$ & $1Lu_{1}$ & $0Ru_{1}$  & $bLu_{1}$ \\
$1$ & $bLu_{2}$ & $1Lu_{2}$  & $0Ru_{3}$  \\
$b$ & $bLu_{3}$ &   & $1Ru_{3}$  \\
\hline 
\end{tabular}
\end{center}
\caption{Table of behaviour for $U_{3,3}$.} \label{tab:U(3,3)}
\end{table}

We give an example of $U_{3,3}$ simulating the three successive Rule 110 timesteps $c_0\vdash c_1\vdash c_2\vdash c_3$ given in Figure~\ref{fig:Rule110Steps}. In the below configurations the current state of $U_{3,3}$ is highlighted in bold, to the left of its tape contents. The tape head position of $U_{3,3}$ is given by an underline and the start state is $u_1$. The configuration immediately below encodes $c_0$ from Figure~\ref{fig:Rule110Steps} with the tape head over cell index~0.
\begin{xalignat*}{7}
&{\pmb{u_1}},\;\ldots\; \tnZero\tnZero\tnOne\tnb&  &\tnZero\tnZero\tnOne\tnb& &\tnZero\tnZero\tnOne\tnb&  &\tnZero\tnZero\tnZero\tnOneUline& &\tnZero\tnb\tnOne\tnOne\tnZero\tnb& &\tnZero\tnb\tnOne\tnOne\tnZero\tnb \ldots\\
\vdash\;\;&{\pmb{u_2}},\;\ldots\; \tnZero\tnZero\tnOne\tnb&  &\tnZero\tnZero\tnOne\tnb&  &\tnZero\tnZero\tnOne\tnb&  &\tnZero\tnZero\tnZeroUline\tnb& &\tnZero\tnb\tnOne\tnOne\tnZero\tnb&  &\tnZero\tnb\tnOne\tnOne\tnZero\tnb \ldots\\
\vdash\;\;&{\pmb{u_1}},\;\ldots\; \tnZero\tnZero\tnOne\tnb&  &\tnZero\tnZero\tnOne\tnb&  &\tnZero\tnZero\tnOne\tnb&   &\tnZero\tnZero\tnZero\tnbUline& &\tnZero\tnb\tnOne\tnOne\tnZero\tnb&   &\tnZero\tnb\tnOne\tnOne\tnZero\tnb \ldots\\
\vdash\;\;&{\pmb{u_3}},\;\ldots\; \tnZero\tnZero\tnOne\tnb&   &\tnZero\tnZero\tnOne\tnb&  &\tnZero\tnZero\tnOne\tnb&   &\tnZero\tnZero\tnZeroUline\tnb&  &\tnZero\tnb\tnOne\tnOne\tnZero\tnb&   &\tnZero\tnb\tnOne\tnOne\tnZero\tnb \ldots\\
\vdash\;\;&{\pmb{u_1}},\;\ldots\; \tnZero\tnZero\tnOne\tnb&   &\tnZero\tnZero\tnOne\tnb& &\tnZero\tnZero\tnOne\tnb&  &\tnZero\tnZeroUline\tnb\tnb&  &\tnZero\tnb\tnOne\tnOne\tnZero\tnb&  &\tnZero\tnb\tnOne\tnOne\tnZero\tnb \ldots\\
\vdash^2\;\;&{\pmb{u_1}},\;\ldots\; \tnZero\tnZero\tnOne\tnb&   &\tnZero\tnZero\tnOne\tnb&  &\tnZero\tnZero\tnOne\tnbUline&   &\tnOne\tnOne\tnb\tnb&  &\tnZero\tnb\tnOne\tnOne\tnZero\tnb&  &\tnZero\tnb\tnOne\tnOne\tnZero\tnb \ldots\\
\vdash\;\;&{\pmb{u_3}},\;\ldots\; \tnZero\tnZero\tnOne\tnb&   &\tnZero\tnZero\tnOne\tnb&  &\tnZero\tnZero\tnOneUline\tnb&   &\tnOne\tnOne\tnb\tnb&  &\tnZero\tnb\tnOne\tnOne\tnZero\tnb&  &\tnZero\tnb\tnOne\tnOne\tnZero\tnb \ldots
\end{xalignat*}
When the tape head reads the subword $1b0$ the left traversal is complete and the right traversal begins.
\begin{xalignat*}{7}
\vdash^6\;\;&{\pmb{u_3}},\;\ldots\; \tnZero\tnZero\tnOne\tnb&  &\tnZero\tnZero\tnOne\tnb& &\tnZero\tnZero\tnZero\tnOne&  &\tnZero\tnZero\tnOne\tnOne&  &\tnZeroUline\tnb\tnOne\tnOne\tnZero\tnb&  &\tnZero\tnb\tnOne\tnOne\tnZero\tnb \ldots\\
\vdash\;\;&{\pmb{u_1}},\;\ldots\; \tnZero\tnZero\tnOne\tnb&   &\tnZero\tnZero\tnOne\tnb& &\mathbf{\tnZero\tnZero\tnZero\tnOne}&   &\mathbf{\tnZero\tnZero\tnOne\tnOneUline}& &\tnb\tnb\tnOne\tnOne\tnZero\tnb&   &\tnZero\tnb\tnOne\tnOne\tnZero\tnb \ldots
\end{xalignat*}
Immediately after the tape head reads a $0$, during a right traversal, the simulation of timestep $c_0\vdash c_1$ is complete. To see this, compare the part of the Turing machine tape in bold with cells $-7$ to $0$ of configuration $c_1$ in Figure~\ref{fig:Rule110Steps}. We continue our simulation to give timestep $c_1\vdash c_2$.
\begin{xalignat*}{7}
\vdash\;\;&{\pmb{u_2}},\;\ldots\; \tnZero\tnZero\tnOne\tnb&   &\tnZero\tnZero\tnOne\tnb&  &\tnZero\tnZero\tnZero\tnOne&  &\tnZero\tnZero\tnOneUline\tnb&  &\tnb\tnb\tnOne\tnOne\tnZero\tnb&  &\tnZero\tnb\tnOne\tnOne\tnZero\tnb \ldots\\
\vdash\;\;&{\pmb{u_2}},\;\ldots\; \tnZero\tnZero\tnOne\tnb&   &\tnZero\tnZero\tnOne\tnb&  &\tnZero\tnZero\tnZero\tnOne&   &\tnZero\tnZeroUline\tnOne\tnb&  &\tnb\tnb\tnOne\tnOne\tnZero\tnb&   &\tnZero\tnb\tnOne\tnOne\tnZero\tnb \ldots\\
\vdash\;\;&{\pmb{u_1}},\;\ldots\; \tnZero\tnZero\tnOne\tnb&   &\tnZero\tnZero\tnOne\tnb&  &\tnZero\tnZero\tnZero\tnOne&  &\tnZero\tnZero\tnOneUline\tnb&  &\tnb\tnb\tnOne\tnOne\tnZero\tnb&   &\tnZero\tnb\tnOne\tnOne\tnZero\tnb \ldots\\
\vdash\;\;&{\pmb{u_2}},\;\ldots\; \tnZero\tnZero\tnOne\tnb&  &\tnZero\tnZero\tnOne\tnb&  &\tnZero\tnZero\tnZero\tnOne&  &\tnZero\tnZeroUline\tnb\tnb&  &\tnb\tnb\tnOne\tnOne\tnZero\tnb& &\tnZero\tnb\tnOne\tnOne\tnZero\tnb \ldots\\
\vdash^3\;\;&{\pmb{u_1}},\;\ldots\; \tnZero\tnZero\tnOne\tnb&  &\tnZero\tnZero\tnOne\tnb&  &\tnZero\tnZero\tnZero\tnOne&  &\tnZeroUline\tnb\tnb\tnb&  &\tnb\tnb\tnOne\tnOne\tnZero\tnb&  &\tnZero\tnb\tnOne\tnOne\tnZero\tnb \ldots\\
\vdash^2\;\;&{\pmb{u_2}},\;\ldots\; \tnZero\tnZero\tnOne\tnb&  &\tnZero\tnZero\tnOne\tnb&  &\tnZero\tnZero\tnZeroUline\tnb&  &\tnOne\tnb\tnb\tnb&  &\tnb\tnb\tnOne\tnOne\tnZero\tnb& &\tnZero\tnb\tnOne\tnOne\tnZero\tnb \ldots\\
\vdash^3\;\;&{\pmb{u_1}},\;\ldots\; \tnZero\tnZero\tnOne\tnb&   &\tnZero\tnZero\tnOne\tnb&  &\tnZero\tnZeroUline\tnb\tnb&  &\tnOne\tnb\tnb\tnb&  &\tnb\tnb\tnOne\tnOne\tnZero\tnb&  &\tnZero\tnb\tnOne\tnOne\tnZero\tnb \ldots\\
\vdash^3\;\;&{\pmb{u_3}},\;\ldots\; \tnZero\tnZero\tnOne\tnb&  &\tnZero\tnZero\tnOneUline\tnb&  &\tnOne\tnOne\tnb\tnb&  &\tnOne\tnb\tnb\tnb&  &\tnb\tnb\tnOne\tnOne\tnZero\tnb&  &\tnZero\tnb\tnOne\tnOne\tnZero\tnb \ldots\\
\vdash^{15}\;\;&{\pmb{u_1}},\;\ldots\; \tnZero\tnZero\tnOne\tnb&  &\mathbf{\tnZero\tnZero\tnZero\tnOne}&  &\mathbf{\tnZero\tnZero\tnOne\tnOne}&  &\mathbf{\tnZero\tnOne\tnOne\tnOne}&  &\mathbf{\tnOne\tnOne\tnZero\tnZeroUline}\tnb\tnb&  &\tnZero\tnb\tnOne\tnOne\tnZero\tnb \ldots
\end{xalignat*}
The simulation of timestep $c_1\vdash c_2$ is complete. To see this, compare the part of the Turing machine tape in bold with cells $-11$ to $4$ of configuration $c_2$ in Figure~\ref{fig:Rule110Steps}. We continue our simulation to give timestep $c_2\vdash c_3$.
\begin{xalignat*}{7}
\vdash^3\;\;&{\pmb{u_2}},\;\ldots\; \tnZero\tnZero\tnOne\tnb&  &\tnZero\tnZero\tnZero\tnOne&  &\tnZero\tnZero\tnOne\tnOne&  &\tnZero\tnOne\tnOne\tnOne&  &\tnOneUline\tnb\tnOne\tnOne\tnb\tnb&  &\tnZero\tnb\tnOne\tnOne\tnZero\tnb \ldots\\
\vdash^4\;\;&{\pmb{u_2}},\;\ldots\; \tnZero\tnZero\tnOne\tnb&  &\tnZero\tnZero\tnZero\tnOne&  &\tnZero\tnZero\tnOne\tnOne&  &\tnZeroUline\tnOne\tnOne\tnOne&  &\tnOne\tnb\tnOne\tnOne\tnb\tnb& &\tnZero\tnb\tnOne\tnOne\tnZero\tnb \ldots\\
\vdash^5\;\;&{\pmb{u_1}},\;\ldots\; \tnZero\tnZero\tnOne\tnb&  &\tnZero\tnZero\tnZero \tnOne&  &\tnZero\tnZero\tnOne\tnOneUline&  &\tnb\tnb\tnOne\tnOne&  &\tnOne\tnb\tnOne\tnOne\tnb\tnb&  &\tnZero\tnb\tnOne\tnOne\tnZero\tnb \ldots\\
\vdash^2\;\;&{\pmb{u_2}},\;\ldots\; \tnZero\tnZero\tnOne\tnb&  &\tnZero\tnZero\tnZero\tnOne&  &\tnZero\tnZeroUline\tnOne\tnb&  &\tnb\tnb\tnOne\tnOne&  &\tnOne\tnb\tnOne\tnOne\tnb\tnb&  &\tnZero\tnb\tnOne\tnOne\tnZero\tnb \ldots\\
\vdash^5\;\;&{\pmb{u_1}},\;\ldots\; \tnZero\tnZero\tnOne\tnb&  &\tnZero\tnZero\tnZero \tnOne&  &\tnZeroUline\tnb\tnb\tnb&  &\tnb\tnb\tnOne\tnOne&  &\tnOne\tnb\tnOne\tnOne\tnb\tnb&  &\tnZero\tnb\tnOne\tnOne\tnZero\tnb \ldots\\
\vdash^2\;\;&{\pmb{u_2}},\;\ldots\; \tnZero\tnZero\tnOne\tnb&  &\tnZero\tnZero\tnZeroUline\tnb&  &\tnOne\tnb\tnb\tnb&  &\tnb\tnb\tnOne\tnOne&  &\tnOne\tnb\tnOne\tnOne\tnb\tnb&  &\tnZero\tnb\tnOne\tnOne\tnZero\tnb \ldots\\
\vdash^6\;\;&{\pmb{u_3}},\;\ldots\; \tnZero\tnZero\tnOneUline\tnb&  &\tnOne\tnOne\tnb\tnb&  &\tnOne\tnb\tnb\tnb&  &\tnb\tnb\tnOne\tnOne&  &\tnOne\tnb\tnOne\tnOne\tnb\tnb&  &\tnZero\tnb\tnOne\tnOne\tnZero\tnb \ldots\\
\vdash^{21}\;\;&{\pmb{u_1}},\;\ldots\; \mathbf{\tnZero\tnZero\tnZero\tnOne}&  &\mathbf{\tnZero\tnZero\tnOne\tnOne}&  &\mathbf{\tnZero\tnOne\tnOne\tnOne}&  &\mathbf{\tnOne\tnOne\tnZero\tnZero}&  &\mathbf{\tnZero\tnOne\tnZero\tnZero\tnOne\tnOneUline}&  &\tnb\tnb\tnOne\tnOne\tnZero\tnb \ldots
\end{xalignat*}
The simulation of timestep $c_2\vdash c_3$ is complete. To see this, compare the part of the Turing machine tape in bold with cells $-15$ to $6$ of configuration $c_3$ in Figure~\ref{fig:Rule110Steps}.

\subsection{$U_{2,4}$}
We begin by describing an \emph{initial} configuration of $U_{2,4}$. To the left of, and including, the tape head position, the Rule 110 state $0$ is encoded by either~$0$ or~$\tnMarkZero$ and the Rule 110 state $1$ is encoded by either~$1$ or~$\tnMarkOne$. The word $\tnMarkZero 1$ is used to terminate a left traversal. To the right of the tape head position, the Rule 110 state $0$ is encoded by $\tnMarkZero$ and the Rule 110 state $1$ is encoded by $\tnMarkOne$ or $0$. The tape symbol $0$ is used to terminate a right traversal. The left and right blank words, from paragraph 4 of Section~\ref{sec:SmallUTMs}, are encoded as $\tnZero\tnZero\tnMarkZero\tnOne$ and $\tnZero\tnMarkOne\tnMarkZero\tnMarkZero\tnZero\tnMarkOne$ respectively. 

\begin{table}[h]
\begin{center}
\begin{tabular}{c@{\;\;}|@{\;\;}c@{\;\;}c}
 & $u_{1}$ & $u_{2}$   \\ \hline
$0$ & $\tnMarkZero Lu_{1}$ & $\tnMarkOne Ru_{1}$  \\
$1$ & $\tnMarkOne Lu_{2}$ & $\tnMarkZero Lu_{2}$  \\
$\tnMarkZero$ & $\tnMarkOne Lu_{1}$ & $0Ru_{2}$  \\
$\tnMarkOne$ & $\tnMarkOne Lu_{1}$ & $1Ru_{2}$  \\
\hline 
\end{tabular}
\end{center}
\caption{\small Table of behaviour for $U_{2,4}$.} \label{tab:U(2,4)}
\end{table}
By way of example we give $U_{2,4}$ simulating the two successive Rule 110 timesteps $c_0\vdash c_1\vdash c_2$ given in Figure~\ref{fig:Rule110Steps}. The configuration immediately below encodes $c_0$ from Figure~\ref{fig:Rule110Steps} with the tape head over cell index~0.
\begin{xalignat*}{7}
&{\pmb{u_1}},\;\ldots\; \tnZero\tnZero\tnMarkZero\tnOne&  &\tnZero\tnZero\tnMarkZero\tnOne& &\tnZero\tnZero\tnMarkZero\tnOne&  &\tnZero\tnZero\tnZero\tnOneUline& &\tnZero\tnMarkOne\tnMarkZero\tnMarkZero\tnZero\tnMarkOne&  &\tnZero\tnMarkOne\tnMarkZero\tnMarkZero\tnZero\tnMarkOne \ldots\\
\vdash^6\;\;&{\pmb{u_1}},\;\ldots\; \tnZero\tnZero\tnMarkZero\tnOne&  &\tnZero\tnZero\tnMarkZero\tnOne& &\tnZero\tnZero\tnMarkZero\tnOneUline& &\tnMarkZero\tnMarkZero\tnMarkOne\tnMarkOne& &\tnZero\tnMarkOne\tnMarkZero\tnMarkZero\tnZero\tnMarkOne& &\tnZero\tnMarkOne\tnMarkZero\tnMarkZero\tnZero\tnMarkOne \ldots\\
\vdash\;\;&{\pmb{u_2}},\;\ldots\; \tnZero\tnZero\tnMarkZero\tnOne&  &\tnZero\tnZero\tnMarkZero\tnOne& &\tnZero\tnZero\tnMarkZeroUline\tnMarkOne&  &\tnMarkZero\tnMarkZero\tnMarkOne\tnMarkOne& &\tnZero\tnMarkOne\tnMarkZero\tnMarkZero\tnZero\tnMarkOne&  &\tnZero\tnMarkOne\tnMarkZero\tnMarkZero\tnZero\tnMarkOne \ldots
\end{xalignat*}
When the tape head reads the subword $\tnMarkZero\tnOne$ the left traversal is complete and the right traversal begins.
\begin{xalignat*}{7}
\vdash^6\;\;&{\pmb{u_2}},\;\ldots\; \tnZero\tnZero\tnMarkZero\tnOne&  &\tnZero\tnZero\tnMarkZero\tnOne&  &\tnZero\tnZero\tnZero\tnOne& &\tnZero\tnZero\tnOne\tnOne&  &\tnZeroUline\tnMarkOne\tnMarkZero\tnMarkZero\tnZero\tnMarkOne& &\tnZero\tnMarkOne\tnMarkZero\tnMarkZero\tnZero\tnMarkOne \ldots\\
\vdash\;\;&{\pmb{u_1}},\;\ldots\; \tnZero\tnZero\tnMarkZero\tnOne&  &\tnZero\tnZero\tnMarkZero\tnOne&  &\mathbf{\tnZero\tnZero\tnZero\tnOne}&   &\mathbf{\tnZero\tnZero\tnOne\tnOne}&   &\tnMarkOne\tnMarkOneUline\tnMarkZero\tnMarkZero\tnZero\tnMarkOne& &\tnZero\tnMarkOne\tnMarkZero\tnMarkZero\tnZero\tnMarkOne \ldots
\end{xalignat*}
Immediately after the tape head reads a $0$, during a right traversal, the simulation of timestep $c_0\vdash c_1$ is complete. To see this, compare the part of the Turing machine tape in bold with cells $-7$ to $0$ of configuration $c_1$ in Figure~\ref{fig:Rule110Steps}. We continue our simulation to give timestep $c_1\vdash c_2$.
\begin{xalignat*}{7}
\vdash^2\;\;&{\pmb{u_1}},\;\ldots\; \tnZero\tnZero\tnMarkZero\tnOne&  &\tnZero\tnZero\tnMarkZero\tnOne&  &\tnZero\tnZero\tnZero\tnOne&   &\tnZero\tnZero\tnOne\tnOneUline&   &\tnMarkOne\tnMarkOne\tnMarkZero\tnMarkZero\tnZero\tnMarkOne& &\tnZero\tnMarkOne\tnMarkZero\tnMarkZero\tnZero\tnMarkOne \ldots\\
\vdash^2\;\;&{\pmb{u_2}},\;\ldots\; \tnZero\tnZero\tnMarkZero\tnOne&  &\tnZero\tnZero\tnMarkZero\tnOne&  &\tnZero\tnZero\tnZero\tnOne&   &\tnZero\tnZeroUline\tnMarkZero\tnMarkOne&   &\tnMarkOne\tnMarkOne\tnMarkZero\tnMarkZero\tnZero\tnMarkOne& &\tnZero\tnMarkOne\tnMarkZero\tnMarkZero\tnZero\tnMarkOne \ldots\\
\vdash\;\;&{\pmb{u_1}},\;\ldots\; \tnZero\tnZero\tnMarkZero\tnOne&  &\tnZero\tnZero\tnMarkZero\tnOne&  &\tnZero\tnZero\tnZero\tnOne&   &\tnZero\tnMarkOne\tnMarkZeroUline\tnMarkOne&   &\tnMarkOne\tnMarkOne\tnMarkZero\tnMarkZero\tnZero\tnMarkOne& &\tnZero\tnMarkOne\tnMarkZero\tnMarkZero\tnZero\tnMarkOne \ldots\\
\vdash^4\;\;&{\pmb{u_2}},\;\ldots\; \tnZero\tnZero\tnMarkZero\tnOne&  &\tnZero\tnZero\tnMarkZero\tnOne&  &\tnZero\tnZero\tnZeroUline\tnMarkOne&   &\tnMarkZero\tnMarkOne\tnMarkOne\tnMarkOne&   &\tnMarkOne\tnMarkOne\tnMarkZero\tnMarkZero\tnZero\tnMarkOne& &\tnZero\tnMarkOne\tnMarkZero\tnMarkZero\tnZero\tnMarkOne \ldots\\
\vdash^5\;\;&{\pmb{u_1}},\;\ldots\; \tnZero\tnZero\tnMarkZero\tnOne&  &\tnZero\tnZero\tnMarkZero\tnOneUline&  &\tnMarkZero\tnMarkZero\tnMarkOne\tnMarkOne&   &\tnMarkZero\tnMarkOne\tnMarkOne\tnMarkOne&   &\tnMarkOne\tnMarkOne\tnMarkZero\tnMarkZero\tnZero\tnMarkOne& &\tnZero\tnMarkOne\tnMarkZero\tnMarkZero\tnZero\tnMarkOne \ldots\\
\vdash\;\;&{\pmb{u_2}},\;\ldots\; \tnZero\tnZero\tnMarkZero\tnOne&  &\tnZero\tnZero\tnMarkZeroUline\tnMarkOne&  &\tnMarkZero\tnMarkZero\tnMarkOne\tnMarkOne&   &\tnMarkZero\tnMarkOne\tnMarkOne\tnMarkOne&   &\tnMarkOne\tnMarkOne\tnMarkZero\tnMarkZero\tnZero\tnMarkOne& &\tnZero\tnMarkOne\tnMarkZero\tnMarkZero\tnZero\tnMarkOne \ldots\\
\vdash^{15}\;\;&{\pmb{u_1}},\;\ldots\; \tnZero\tnZero\tnMarkZero\tnOne&  &\mathbf{\tnZero\tnZero\tnZero\tnOne}&  &\mathbf{\tnZero\tnZero\tnOne\tnOne}&   &\mathbf{\tnZero\tnOne\tnOne\tnOne}&   &\mathbf{\tnOne\tnOne\tnZero\tnZero}\tnMarkOne\tnMarkOneUline& &\tnZero\tnMarkOne\tnMarkZero\tnMarkZero\tnZero\tnMarkOne \ldots
\end{xalignat*}
The simulation of timestep $c_1\vdash c_2$ is complete. To see this, compare the part of the Turing machine tape in bold with cells $-11$ to $4$ of configuration $c_2$ in Figure~\ref{fig:Rule110Steps}.

\subsection{$U_{6,2}$}
We begin by describing an \emph{initial} configuration of $U_{6,2}$. To the left of, and including, the tape head position, the Rule 110 state $0$ is encoded by the word $00$ and the Rule 110 state $1$ is encoded by the word $11$. The word $010100$ is used to terminate a left traversal and encodes the sequence of Rule 110 states $010$. To the right of the tape head position the Rule 110 state $0$ is encoded by the word~$00$ and the Rule 110 state $1$ is encoded by either of the words $01$ or $10$. The word~$10$ is used to terminate a right traversal. The left and right blank words, from paragraph~4 of Section~\ref{sec:SmallUTMs}, are encoded as $\tnZero\tnZero\tnZero\tnZero\tnZero\tnOne\tnZero\tnOne$ and $\tnOne\tnZero \tnZero\tnOne \tnZero\tnZero\tnZero\tnZero \tnOne\tnZero \tnZero\tnOne$ respectively. 

\begin{table}[h]
\begin{center}
\begin{tabular}{c@{\;\;}|@{\;\;}c@{\;\;}c@{\;\;}c@{\;\;}c@{\;\;}c@{\;\;}c}
 & $u_{1}$ & $u_{2}$  & $u_{3}$ & $u_{4}$  & $u_{5}$ & $u_{6}$ \\ \hline
$0$ & $0Lu_{1}$ & $0Lu_{6}$  & $0Ru_{2}$ & $1Ru_{5}$  & $1Lu_{4}$ & $1Lu_{1}$\\
$1$ & $1Lu_{2}$ & $0Lu_{3}$  & $1Lu_{3}$ & $0Ru_{6}$  & $1Ru_{4}$ & $0Ru_{4}$ \\
\hline 
\end{tabular}
\end{center}
\caption{\small Table of behaviour for $U_{6,2}$.} \label{tab:U(6,2)}
\end{table}

To illustrate the operation of $U_{6,2}$ we simulate the Rule 110 timestep $c_0\vdash c_1$ given in Figure~\ref{fig:Rule110Steps}. The configuration immediately below encodes $c_0$ from Figure~\ref{fig:Rule110Steps} with the tape head over cell index~0.
\begin{xalignat*}{4}
&{\pmb{u_1}},\;\ldots\tnZero\tnZero\tnZero\tnZero\tnZero\tnOne\tnZero\tnOne& &\tnZero\tnZero\tnZero\tnZero\tnZero\tnOne\tnZero\tnOne&  &\tnZero\tnZero\tnZero\tnZero\tnZero\tnZero\tnOne\tnOneUline& &\tnOne\tnZero \tnZero\tnOne \tnZero\tnZero\tnZero\tnZero \tnOne\tnZero \tnZero\tnOne\ldots\\
\vdash\;\;&{\pmb{u_2}},\;\ldots\tnZero\tnZero\tnZero\tnZero\tnZero\tnOne\tnZero\tnOne& &\tnZero\tnZero\tnZero\tnZero\tnZero\tnOne\tnZero\tnOne&  &\tnZero\tnZero\tnZero\tnZero\tnZero\tnZero\tnOneUline\tnOne& &\tnOne\tnZero \tnZero\tnOne \tnZero\tnZero\tnZero\tnZero \tnOne\tnZero \tnZero\tnOne\ldots\\
\vdash\;\;&{\pmb{u_3}},\;\ldots\tnZero\tnZero\tnZero\tnZero\tnZero\tnOne\tnZero\tnOne& &\tnZero\tnZero\tnZero\tnZero\tnZero\tnOne\tnZero\tnOne&  &\tnZero\tnZero\tnZero\tnZero\tnZero\tnZeroUline\tnZero\tnOne& &\tnOne\tnZero \tnZero\tnOne \tnZero\tnZero\tnZero\tnZero \tnOne\tnZero \tnZero\tnOne\ldots\\
\vdash\;\;&{\pmb{u_2}},\;\ldots\tnZero\tnZero\tnZero\tnZero\tnZero\tnOne\tnZero\tnOne& &\tnZero\tnZero\tnZero\tnZero\tnZero\tnOne\tnZero\tnOne&  &\tnZero\tnZero\tnZero\tnZero\tnZero\tnZero\tnZeroUline\tnOne& &\tnOne\tnZero \tnZero\tnOne \tnZero\tnZero\tnZero\tnZero \tnOne\tnZero \tnZero\tnOne\ldots\\
\vdash\;\;&{\pmb{u_6}},\;\ldots\tnZero\tnZero\tnZero\tnZero\tnZero\tnOne\tnZero\tnOne& &\tnZero\tnZero\tnZero\tnZero\tnZero\tnOne\tnZero\tnOne&  &\tnZero\tnZero\tnZero\tnZero\tnZero\tnZeroUline\tnZero\tnOne& &\tnOne\tnZero \tnZero\tnOne \tnZero\tnZero\tnZero\tnZero \tnOne\tnZero \tnZero\tnOne\ldots\\
\vdash\;\;&{\pmb{u_1}},\;\ldots\tnZero\tnZero\tnZero\tnZero\tnZero\tnOne\tnZero\tnOne& &\tnZero\tnZero\tnZero\tnZero\tnZero\tnOne\tnZero\tnOne&  &\tnZero\tnZero\tnZero\tnZero\tnZeroUline\tnOne\tnZero\tnOne& &\tnOne\tnZero \tnZero\tnOne \tnZero\tnZero\tnZero\tnZero \tnOne\tnZero \tnZero\tnOne\ldots\\
\vdash^5\;\;&{\pmb{u_1}},\;\ldots\tnZero\tnZero\tnZero\tnZero\tnZero\tnOne\tnZero\tnOne& &\tnZero\tnZero\tnZero\tnZero\tnZero\tnOne\tnZero\tnOneUline&  &\tnZero\tnZero\tnZero\tnZero\tnZero\tnOne\tnZero\tnOne& &\tnOne\tnZero \tnZero\tnOne \tnZero\tnZero\tnZero\tnZero \tnOne\tnZero \tnZero\tnOne\ldots\\
\vdash\;\;&{\pmb{u_2}},\;\ldots\tnZero\tnZero\tnZero\tnZero\tnZero\tnOne\tnZero\tnOne& &\tnZero\tnZero\tnZero\tnZero\tnZero\tnOne\tnZeroUline\tnOne&  &\tnZero\tnZero\tnZero\tnZero\tnZero\tnOne\tnZero\tnOne& &\tnOne\tnZero \tnZero\tnOne \tnZero\tnZero\tnZero\tnZero \tnOne\tnZero \tnZero\tnOne\ldots\\
\vdash\;\;&{\pmb{u_6}},\;\ldots\tnZero\tnZero\tnZero\tnZero\tnZero\tnOne\tnZero\tnOne& &\tnZero\tnZero\tnZero\tnZero\tnZero\tnOneUline\tnZero\tnOne&  &\tnZero\tnZero\tnZero\tnZero\tnZero\tnOne\tnZero\tnOne& &\tnOne\tnZero \tnZero\tnOne \tnZero\tnZero\tnZero\tnZero \tnOne\tnZero \tnZero\tnOne\ldots\\
\vdash\;\;&{\pmb{u_4}},\;\ldots\tnZero\tnZero\tnZero\tnZero\tnZero\tnOne\tnZero\tnOne& &\tnZero\tnZero\tnZero\tnZero\tnZero\tnZero\tnZeroUline\tnOne&  &\tnZero\tnZero\tnZero\tnZero\tnZero\tnOne\tnZero\tnOne& &\tnOne\tnZero \tnZero\tnOne \tnZero\tnZero\tnZero\tnZero \tnOne\tnZero \tnZero\tnOne\ldots
\end{xalignat*}
When the tape head reads the subword $\tnOne\tnZero\tnOne\tnZero\tnZero$ the left traversal is complete and the right traversal begins.

\begin{xalignat*}{4}
\vdash\;\;&{\pmb{u_5}},\;\ldots\tnZero\tnZero\tnZero\tnZero\tnZero\tnOne\tnZero\tnOne& &\tnZero\tnZero\tnZero\tnZero\tnZero\tnZero\tnOne\tnOneUline&  &\tnZero\tnZero\tnZero\tnZero\tnZero\tnOne\tnZero\tnOne& &\tnOne\tnZero \tnZero\tnOne \tnZero\tnZero\tnZero\tnZero \tnOne\tnZero \tnZero\tnOne\ldots\\
\vdash\;\;&{\pmb{u_4}},\;\ldots\tnZero\tnZero\tnZero\tnZero\tnZero\tnOne\tnZero\tnOne& &\tnZero\tnZero\tnZero\tnZero\tnZero\tnZero\tnOne\tnOne&  &\tnZeroUline\tnZero\tnZero\tnZero\tnZero\tnOne\tnZero\tnOne& &\tnOne\tnZero \tnZero\tnOne \tnZero\tnZero\tnZero\tnZero \tnOne\tnZero \tnZero\tnOne\ldots\\
\vdash\;\;&{\pmb{u_5}},\;\ldots\tnZero\tnZero\tnZero\tnZero\tnZero\tnOne\tnZero\tnOne& &\tnZero\tnZero\tnZero\tnZero\tnZero\tnZero\tnOne\tnOne&  &\tnOne\tnZeroUline\tnZero\tnZero\tnZero\tnOne\tnZero\tnOne& &\tnOne\tnZero \tnZero\tnOne \tnZero\tnZero\tnZero\tnZero \tnOne\tnZero \tnZero\tnOne\ldots\\
\vdash\;\;&{\pmb{u_4}},\;\ldots\tnZero\tnZero\tnZero\tnZero\tnZero\tnOne\tnZero\tnOne& &\tnZero\tnZero\tnZero\tnZero\tnZero\tnZero\tnOne\tnOne&  &\tnOneUline\tnOne\tnZero\tnZero\tnZero\tnOne\tnZero\tnOne& &\tnOne\tnZero \tnZero\tnOne \tnZero\tnZero\tnZero\tnZero \tnOne\tnZero \tnZero\tnOne\ldots\\
\vdash\;\;&{\pmb{u_6}},\;\ldots\tnZero\tnZero\tnZero\tnZero\tnZero\tnOne\tnZero\tnOne& &\tnZero\tnZero\tnZero\tnZero\tnZero\tnZero\tnOne\tnOne&  &\tnZero\tnOneUline\tnZero\tnZero\tnZero\tnOne\tnZero\tnOne& &\tnOne\tnZero \tnZero\tnOne \tnZero\tnZero\tnZero\tnZero \tnOne\tnZero \tnZero\tnOne\ldots\\
\vdash\;\;&{\pmb{u_4}},\;\ldots\tnZero\tnZero\tnZero\tnZero\tnZero\tnOne\tnZero\tnOne& &\tnZero\tnZero\tnZero\tnZero\tnZero\tnZero\tnOne\tnOne&  &\tnZero\tnZero\tnZeroUline\tnZero\tnZero\tnOne\tnZero\tnOne& &\tnOne\tnZero \tnZero\tnOne \tnZero\tnZero\tnZero\tnZero \tnOne\tnZero \tnZero\tnOne\ldots\\
\vdash^4\;\;&{\pmb{u_4}},\;\ldots\tnZero\tnZero\tnZero\tnZero\tnZero\tnOne\tnZero\tnOne& &\tnZero\tnZero\tnZero\tnZero\tnZero\tnZero\tnOne\tnOne&  &\tnZero\tnZero\tnZero\tnZero\tnZeroUline\tnOne\tnZero\tnOne& &\tnOne\tnZero \tnZero\tnOne \tnZero\tnZero\tnZero\tnZero \tnOne\tnZero \tnZero\tnOne\ldots\\
\vdash\;\;&{\pmb{u_5}},\;\ldots\tnZero\tnZero\tnZero\tnZero\tnZero\tnOne\tnZero\tnOne& &\tnZero\tnZero\tnZero\tnZero\tnZero\tnZero\tnOne\tnOne&  &\tnZero\tnZero\tnZero\tnZero\tnOne\tnOneUline\tnZero\tnOne& &\tnOne\tnZero \tnZero\tnOne \tnZero\tnZero\tnZero\tnZero \tnOne\tnZero \tnZero\tnOne\ldots\\
\vdash\;\;&{\pmb{u_4}},\;\ldots\tnZero\tnZero\tnZero\tnZero\tnZero\tnOne\tnZero\tnOne& &\tnZero\tnZero\tnZero\tnZero\tnZero\tnZero\tnOne\tnOne&  &\tnZero\tnZero\tnZero\tnZero\tnOne\tnOne\tnZeroUline\tnOne& &\tnOne\tnZero \tnZero\tnOne \tnZero\tnZero\tnZero\tnZero \tnOne\tnZero \tnZero\tnOne\ldots\\
\vdash^2\;\;&{\pmb{u_4}},\;\ldots\tnZero\tnZero\tnZero\tnZero\tnZero\tnOne\tnZero\tnOne& &\tnZero\tnZero\tnZero\tnZero\tnZero\tnZero\tnOne\tnOne&  &\tnZero\tnZero\tnZero\tnZero\tnOne\tnOne\tnOne\tnOne& &\tnOneUline\tnZero \tnZero\tnOne \tnZero\tnZero\tnZero\tnZero \tnOne\tnZero \tnZero\tnOne\ldots\\
\vdash\;\;&{\pmb{u_6}},\;\ldots\tnZero\tnZero\tnZero\tnZero\tnZero\tnOne\tnZero\tnOne& &\tnZero\tnZero\tnZero\tnZero\tnZero\tnZero\tnOne\tnOne&  &\tnZero\tnZero\tnZero\tnZero\tnOne\tnOne\tnOne\tnOne& &\tnZero\tnZeroUline \tnZero\tnOne \tnZero\tnZero\tnZero\tnZero \tnOne\tnZero \tnZero\tnOne\ldots\\
\vdash\;\;&{\pmb{u_1}},\;\ldots\tnZero\tnZero\tnZero\tnZero\tnZero\tnOne\tnZero\tnOne& &\mathbf{\tnZero\tnZero\tnZero\tnZero\tnZero\tnZero\tnOne\tnOne}&  &\mathbf{\tnZero\tnZero\tnZero\tnZero\tnOne\tnOne\tnOne\tnOne}& &\tnZeroUline\tnOne \tnZero\tnOne \tnZero\tnZero\tnZero\tnZero \tnOne\tnZero \tnZero\tnOne\ldots
\end{xalignat*}
Immediately after the tape head reads a $10$, during a right traversal, the simulation of timestep $c_0\vdash c_1$ is complete. To see this, compare the part of the Turing machine tape in bold (recall $0$ and $1$ are encoded as $00$ and $11$ respectively) with cells $-7$ to $0$ of configuration $c_1$ in Figure~\ref{fig:Rule110Steps}.

\section{Discussion on lower bounds}
The pursuit to find the smallest possible universal Turing machine must also involve the search for lower bounds, finding the largest set of Turing machines that are in some sense non-universal. One approach is to settle the decidability of the halting problem, but this approach is not suitable for the machines we have presented.

It is known that the halting problem is decidable for (standard) Turing machines with the following state-symbol pairs $(2,2)$~\cite{Kudlek1996,Pavlotskaya1973}, $(3,2)$~\cite{Pavlotskaya1978}, $(2,3)$~(claimed by Pavlotskaya~\cite{Pavlotskaya1973}), $(1,n)$~\cite{Hermann1968} and $(n,1)$ (trivial), where $n\geqslant 1$. Then, these decidability results imply that a universal Turing machine, that simulates any Turing machine $M$ and halts if and only if $M$ halts, is not possible for these state-symbol pairs. Hence these results give lower bounds on the size of universal machines of this type. While it is trivial to prove that the halting problem is decidable for (possibly halting) weak machines with state-symbol pairs of the form $(n,1)$, it is not known whether the above decidability results generalise to (possibly halting) weak Turing machines.

The weak machines presented in this paper, and those in~\cite{Cook2004,Wolfram2002}, do not halt. Hence the non-universality results discussed in the previous paragraph would have to be generalised to non-halting weak machines to give lower bounds that are relevant for our machines. This may prove difficult for two reasons. The first issue is that, intuitively speaking, weakness gives quite an advantage. For instance, the program of a universal machine may be encoded in one of the infinitely repeated blank words of the weak machine. The second issue is related to the problem of defining a computation. Informally, a computation could be defined as a sequence of configurations that leads to a special terminal configuration. For non-halting machines, there are many ways to define a terminal configuration. Given a definition of terminal configuration we may prove that the terminal configuration problem (will a machine ever enter a terminal configuration) is decidable for a machine or set of machines. However this result may not hold as a proof of non-universality if we subsequently alter our definition of terminal configuration.

\section*{Acknowledgements}
Turlough Neary is funded by the Irish Research Council for Science, Engineering and Technology. Damien Woods is funded by Science Foundation Ireland grant number 04/IN3/1524.

\bibliographystyle{plain}
\bibliography{NearyWoods-Small_Weak_UTMs}
\end{document}